\documentclass[fleqn,usenatbib]{mnras}

\usepackage{newtxtext,newtxmath}

\usepackage[T1]{fontenc}
\DeclareRobustCommand{\VAN}[3]{#2}
\let\VANthebibliography\thebibliography
\def\thebibliography{\DeclareRobustCommand{\VAN}[3]{##3}\VANthebibliography}

\usepackage{graphicx}	
\usepackage{amsmath}	

\title[Extragalactic GC Origins from $\text{[Eu/}\alpha\text{]}$]{On the feasibility of using $\text{[Eu/}\alpha\text{]}$ to identify accreted and in-situ extragalactic globular clusters
}

\author[A. Gvozdenko et al.]{
A. Gvozdenko,$^{1}$\thanks{E-mail: anastasia.gvozdenko@durham.ac.uk}
M. A. Beasley,$^{2,3,4}$
K. Fahrion,$^{5}$ 
S. Larsen,$^{6}$ 
F. La Barbera,$^{7}$ 
A. Vazdekis$^{2,3}$ 
\newauthor
and I. Mart\'{\i}n-Navarro$^{2,3}$
\\
$^{1}$Department of Physics, Centre for Extragalactic Astronomy, Durham University, South Road, Durham DH1 3LE, UK\\
$^{2}$Instituto de Astrofísica de Canarias, Calle V\'{i}a L\'{a}ctea, E-38206 La Laguna, Spain.\\
$^{3}$Departamento de Astrof\'{i}sica, Universidad de La Laguna, E-38206 La Laguna, Spain\\
$^{4}$Centre for Astrophysics and Supercomputing, Swinburne University, John Street, Hawthorn VIC 3122, Australia \\
$^{5}$Department of Astrophysics, University of Vienna, T\"{u}rkenschanzstra{\ss}e 17, 1180 Wien, Austria \\
$^{6}$Department of Astrophysics/IMAPP, Radboud University, PO Box 9010, 6500 GL Nijmegen, The Netherlands \\
$^{7}$INAF-Osservatorio Astronomico di Capodimonte sal. Moiariello 16 Napoli 80131, Italy 
}

\date{Accepted XXX. Received YYY; in original form ZZZ}

\pubyear{\the\year{}}

\begin{document}
\label{firstpage}
\pagerange{\pageref{firstpage}--\pageref{lastpage}}
\maketitle

\begin{abstract}

The chemical abundance ratio [Eu/$\alpha$] has been proposed as a discriminator between ex-situ and in-situ stellar populations in the Milky Way and external galaxies. Here we show with a high-quality, homogeneous dataset of integrated-light abundances for 28 globular clusters (GCs), where 21 are extragalactic and 7 galactic, that this behaviour holds in other Local Group galaxies. In particular, we find metal-rich ([Fe/H] > --0.7) M31 GCs have uniformly low [Eu/Si]$\sim-0.4$, indicating they represent an in-situ population. 
Lower metallicity GCs in M31 generally show high [Eu/Si] $\sim+0.4$, suggesting they originated in accreted progenitors, and these GCs lie at larger projected radii from the M31 centre. 
Furthermore, we compare the M31 GC abundances with similar determinations for the M31 bulge for Na, Mg, Si, Ti, Ca, Eu, Cr, Mn, Ba, Ni, and Cu. The in-situ M31 GCs and M31 bulge exhibit similar abundance trends, excepting Ba -- and possibly Cr and Ti, suggesting similar star formation processes.  [Eu/Si] in the M31 bulge is consistent with in-situ Milky Way stars, but is higher than the M31 in-situ GCs.
For M33 and intermediate-mass dwarf galaxies, we identify a mixture of in-situ and ex-situ GCs, while for lower-mass dwarfs, we identify primarily in-situ GCs.
These results suggest that accreted GCs in massive galaxies originated, at least in part, from intermediate-mass, intermediate-metallicity progenitors.
We support earlier suggestions that [Eu/$\alpha$] is a powerful in-/ex-situ separator for extragalactic GCs, demonstrating the potential of high-resolution integrated-light abundances for extragalactic archaeology.
\end{abstract}

\begin{keywords}
Galaxy: globular clusters: general –
galaxies: abundances –
galaxies: formation –
galaxies: evolution –
galaxies: Local Group
\end{keywords}



\section{Introduction}
 
The assembly history of galaxies can be traced by the processes through which they accumulate their stellar populations. In particular, since star formation timescales and chemical enrichment are a function of host galaxy mass, the stellar population properties of the progenitor (in-situ) and accreted (ex-situ) components of galaxies are expected to differ (e.g., \citealt{Tinsley1979, Matteucci1986, Wheeler1989, Freeman2002}). 
Globular clusters (GCs) have long been used as bright, surviving tracers of these accretion events.

In the Milky Way (MW), kinematic and chemical tagging of stars and GCs has revealed common orbital and abundance signatures that link them to accreted dwarf galaxies (e.g., \citealt{Nissen2010, Buder2021, Dodd2023}).
These accreted populations often exhibit specific dynamical and chemical characteristics that stand out against stars formed within the in situ population \citep{RecioBlanco2021}. Specifically, it has been shown that accreted halo stars at high orbital energy exhibit elevated ratios of \textit{r}-process to $\alpha$-elements, e.g. [Eu/Si] (or sometimes [Eu/Ca]), whereas for in-situ stars at lower orbital energies these abundance ratios are lower \citep{Monty2024}. 
The general interpretation of the behaviour of [Eu/Si] is that $\alpha$-elements are primarily produced in core-collapse supernovae on short timescales, while a major site for the production of \textit{r}-process elements is thought to be neutron-star mergers, which presumably operate on much longer timescales. Furthermore, \citet{Kobayashi2023} indicate that Eu enrichment cannot be explained by neutron-star mergers alone, but instead requires contributions from both prompt and delayed \textit{r}-process sources. Consequently, elevated [Eu/$\alpha$] ratios are expected in systems that experienced more extended star formation histories compared to those formed through rapid, short-duration bursts. Therefore, the ratio between these two classes of elements can act as a "cosmic clock". In this picture, a high \textit{r}-process/$\alpha$-element ratio reflects the longer star formation timescales expected in the progenitors of accreted populations, which are typically lower-mass systems that were subsequently incorporated into more massive host galaxies (e.g., \citealt{Hirai2025}).

Thus, the ratio of \textit{r}-process to $\alpha$ elements provides a powerful chemical tag to trace accretion histories across a wide range of metallicities. 
By compiling a comprehensive dataset of abundances for 54 MW GCs, \citet{Monty2024} found that the systematic differences in [Eu/Si] observed in field stars extend consistently to GC populations. Specifically, at high orbital energy, where accretion-dominated populations are found, GCs exhibit elevated [Eu/Si] values, while GCs at lower energies, predominantly formed in-situ, display lower levels. This consistency implies that GCs trace the star formation histories of their host galaxies, making them valuable tools to separate the accreted and in-situ populations within the same GC system. Furthermore, the utility of [Eu/Si] as a chemical tag for linking GCs to their progenitor galaxies may extend beyond the MW and Local Group. 

High-resolution cosmological simulations demonstrate that Eu can be employed as an effective chemical tag to distinguish accreted stellar components from those formed within the massive galaxies. \citet{Hirai2025} found that in a simulated Milky Way-like galaxy, accreted stars, formed in low-mass dwarf galaxies, exhibit larger scatter and higher [Eu/Fe] ratios due to enrichment from rare neutron star mergers, whereas in-situ stars show more uniform and lower [Eu/Fe] ratios owing to well-mixed chemical evolution in the galactic disc. 

In this paper, we further explore the idea suggested in \citet{Monty2024} and extend the analysis to extragalactic GC systems. We use the results from \citet{Larsen2022} and \citet{Gvozdenko2024} to construct a homogeneous sample of 46 Local Group GCs, of which 21 extragalactic and 7 MW GCs have both Eu and Si abundance measurements. This sample allows us to examine high-quality integrated-light abundances for GCs in MW, M31, M33, N2403, N205, N6822, N185, Sextans~A, Fornax, and WLM. Furthermore, for the M31 system, we draw a comparison between its GCs and the bulge abundances derived from integrated light spectra using methods from \citet{LaBarbera2024, LaBarbera2026}.

In what follows, we describe the data in Section~\ref{sec:data}. In  Section~\ref{sec:Results} we focus on the chemistry and the general abundance trends of the sample. A detailed discussion on MW, M31 and dwarf GC systems and a comparison to the M31 bulge findings is given in Section~\ref{sec:disc}. We present our conclusions in Section~\ref{sec:concs}.

\section{Data}
\label{sec:data}

A key aspect of this analysis is the homogeneous nature of the adopted data. Integrated light abundances for 45 GCs were taken from \cite{Larsen2022} with an additional GC in the Sextans~A dwarf galaxy from \cite{Gvozdenko2024}. 
The data was observed with high-resolution spectrographs using the HIRES spectrograph on the Keck I telescope for northern targets and UVES on the ESO Very Large Telescope for southern targets.
In total, 46 GCs have been studied in \citep{Larsen2022, Gvozdenko2024}, to derive direct measurements of  Na, Mg, Si, Ca, Sc, Ti, Cr, Mn, Fe, Ni, Cu, Zn, Zr, Ba, Eu  (see Section~\ref{sec:Abundances}). For 28 (21 extragalactic and 7 galactic), both values of Si and Eu were successfully derived; the list of these targets is given in Table\,\ref{tab:lg_gc_eus_fe}. The same analysis methods and models were used in both of the above studies.

A detailed discussion of the procedures followed to measure abundances from the integrated light spectra of GCs is given in \cite{Larsen2022} (also \citealt{Larsen2012, Larsen2017, Larsen2018}).
Briefly, the analysis involves computing simple stellar population (SSP) model spectra and adjusting individual element abundances to achieve the best fits to the observed spectrum using theoretical atmospheres (via a Python wrapper iSpy3\footnote{https://github.com/soerenslarsen/ISPy3}). For cool stars (Teff$<$4000 K) MARCS and TURBOSPECTRUM are used \citep{Plez2012}, and for hotter stars, ATLAS12 models are used in combination with SYNTHE \citep{Kurucz1979}. The studies incorporated corrections for non-local thermodynamic equilibrium (NLTE) effects for the elements Na, Mg, Ca, Ti, Mn, Fe, Ni, and Ba based on an approach tailored for SSP modelling \citep{Eitner2019}. For one of the targets, Sextans~A GC-1, only the upper limit for Eu was found. 
Namely, while the fit converged towards a best-fit value, no lower bound on the uncertainty range was found (i.e. the data are consistent with the complete absence of the line), and hence only a positive error bar is defined.

We include additional abundance estimates for the M31 bulge from \citet{LaBarbera2026}, who analysed long-slit integrated-light spectra with the E-MILES \citep{Vazdekis2016} and \citet{Conroy2018} stellar population models, using the 
response functions to capture the impact of varying elemental abundance ratios. For Na, Mg, Si, Ca, and Ti, we used the radial abundance measurements published by \citet{LaBarbera2026}.

We also constructed a stacked spectrum by averaging the \citet{LaBarbera2026} data over galactocentric radii of 200–600 pc, and derived abundances from this spectrum for Na, Mg, Si, Ca, and Ti, as well as for the additional elements examined here: Cr, Mn, Ba, Ni, Cu, and Eu. We followed the \citet{LaBarbera2026} procedure, but relied exclusively on full spectral fitting (instead of combining spectral- and index-based constraints), because an index analysis for several species (e.g. Eu, Cr, Mn) would require a dedicated characterisation of the relevant absorption features, which is beyond the scope of the present work.
The description of the methods and all of the later plotted abundance measurements are provided in Appendix\,\ref{apx:M31bulge}.

\section{Results}
\label{sec:Results}

\subsection{Element abundances}
\label{sec:Abundances}

\begin{figure}
    \centering
    \includegraphics[width=8.5cm]{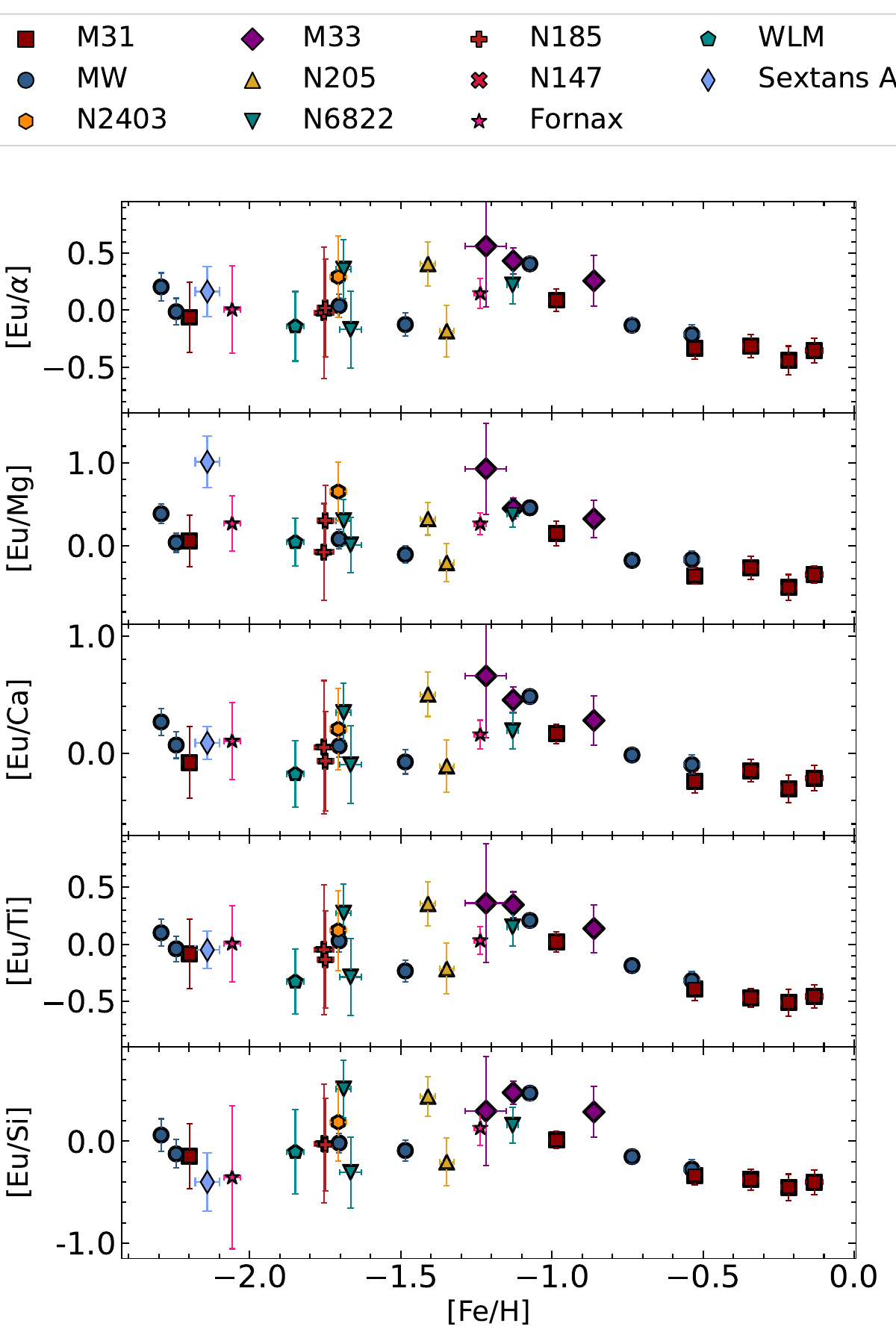}
    \caption{Element abundance ratios [Eu/$\alpha$] (top), [Eu/Mg] (second), [Eu/Ca] (third), [Eu/Ti] (fourth), and Eu/Si (bottom) plotted against [Fe/H] for the Local Group GCs.}
    \label{fig:r-versus_alpha}
\end{figure}

To study the GC systems' formation histories in ways similar to those proposed by \citet{Monty2024}, we explore the relations of Eu with the $\alpha$-elements Mg, Ca, and Ti, as well as with an average-$\alpha$ abundance (the mean of Mg, Si, Ca, and Ti), in Figure~\ref{fig:r-versus_alpha}. Unsurprisingly, all four panels show behaviour very similar to [Eu/Si], albeit with slightly larger scatter at the metal-poor end. Since Mg and Ca are relatively easy to measure in integrated-light spectra, these elements offer promising alternatives to Si for this sort of analysis. However, we highlight the caveat that some MW GCs show significant internal spreads in Mg (similar to the well-known Na spreads; \citealt{Carretta2009}), which may complicate their interpretation. In addition, some Local Group GCs appear highly depleted in Mg at low metallicities -- beyond what might be expected from internal abundance spreads -- leading to anomalously high [Eu/Mg] (e.g., Sextans~A GC-1\footnote{Only an upper limit on its [Eu/Fe] value was found; all upper limits appear consistent with the current picture of the abundance pattern of Local Group GCs.}). Therefore, for our available sample we could equally use Ca or Si; however, \citet{Monty2024} suggest that Si is the "purer" of the two 
$\alpha$-elements, and indeed use [Eu/Ca] interchangeably with [Eu/Si] only when Si values are not available. For the data set presented here, however, Si and Ca are always treated separately, since both were derived for the same set of GCs in our sample (the same 28 out of the full sample of 46).

We therefore focus our attention on the abundance ratios involving Eu and Si, with [Eu/Si] serving as our primary diagnostic. Eu is regarded as a predominantly \textit{r}-process element (e.g. \citealt{Burris2000, Prantzos2020}); solar-system studies indicate that Eu is 94--97 per cent \textit{r}-process, with only a minor contribution from the \textit{s}-process \citep{Travaglio1999}. It is also the only readily measurable \textit{r}-process element in the optical range for integrated-light spectra of GCs. 
Si can be considered a good tracer of core-collapse SN nucleosynthesis in GCs, since it shows minimal internal spreads in MW GCs (unlike, e.g., Mg at low metallicities).

\subsection{Classification of the Local group GC based on [Eu/Si] to [Fe/H] relation}

\begin{figure*}
    \centering
    \includegraphics[width=12cm]{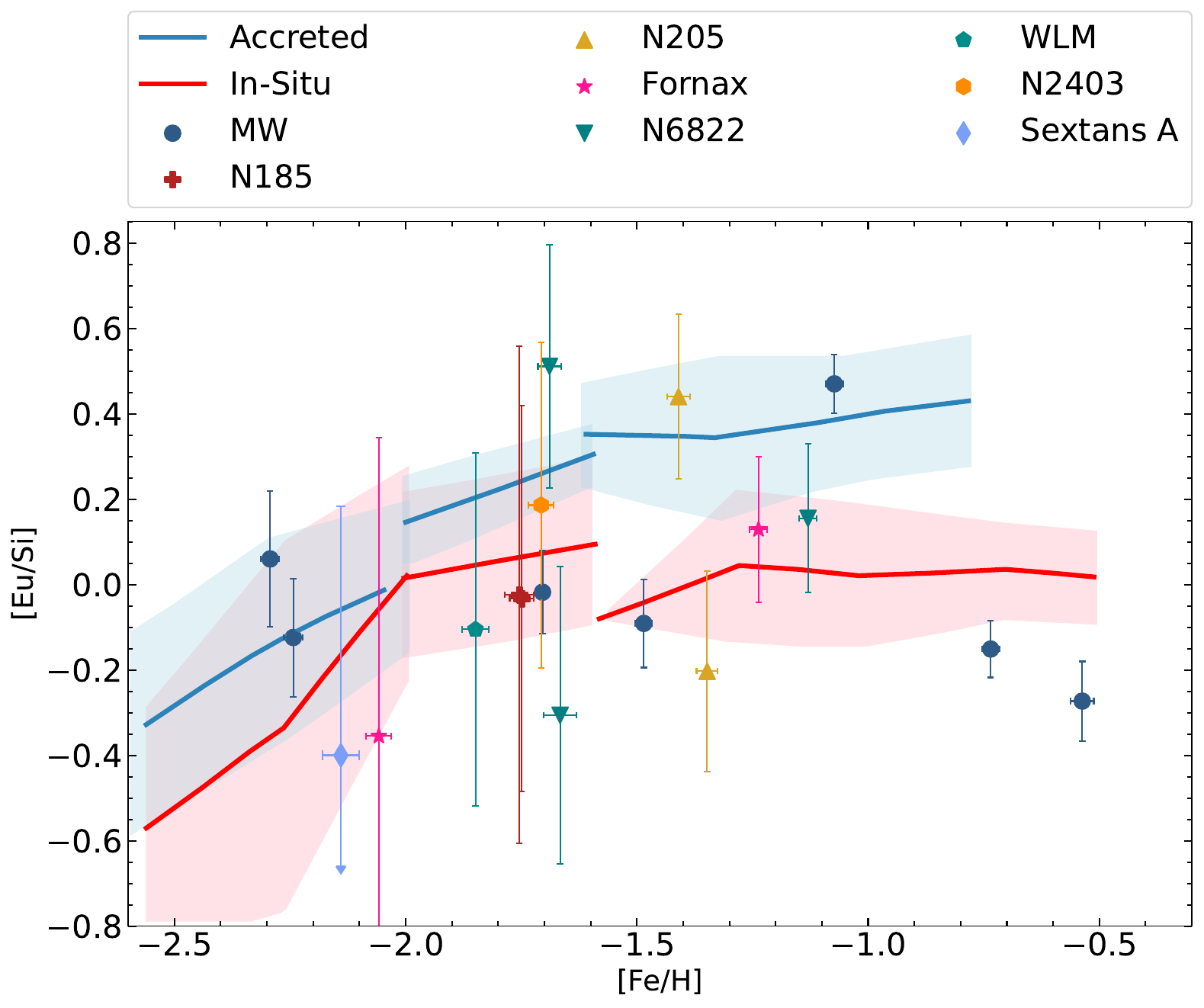}
    \caption{[Eu/Si] plotted against [Fe/H] for the GCs in the MW and Local Group dwarf galaxies. The red and blue lines are defined for in-situ and ex-situ stellar populations taken from \citet{Monty2024}. The shaded regions around the lines represent the 1$\sigma$ spread in [Eu/Si].}
    \label{fig:Eu_fe_GC-MW-dwarf}
\end{figure*}

The [Eu/Si]--[Fe/H] relation defined by \citet{Monty2024} was derived using MW stellar populations and is therefore calibrated for the chemical evolution of the MW. Nevertheless, we investigate whether this empirical relation can be extended to GC systems in other Local Group galaxies. Figure~\ref{fig:Eu_fe_GC-MW-dwarf} shows the [Eu/Si]--[Fe/H] relation for the GCs in the Local Group dwarf galaxies and the MW, with the in-situ and accreted trends adopted from the analysis of \citet{Monty2024}. The red and blue relations correspond to the in-situ and accreted stellar populations, respectively. At intermediate metallicities ($-1.6<$[Fe/H]$<-1.0$), the data suggest a potential mixture of in-situ / ex-situ GCs in host galaxies. At the lowest metallicities  ([Fe/H]$<-1.6$) the discriminating power in [Eu/Si] seems to become increasingly lost as the two sequences converge. Although there may be an intrinsic spread in [Eu/Si] at these metallicities, the current data do not permit a more detailed assessment.

Figure~\ref{fig:Eu_fe_GC-massive-gal} extends this comparison to the more massive Local Group galaxies, M31 and M33. The M31 GCs extend the observed metallicity range beyond that sampled by the MW calibration. At higher metallicities ($\mathrm{[Fe/H]} > -1.0$), the M31 and MW GCs form a tightly clustered sequence lying at or below the in-situ relation ([Eu/Si] $< 0$).
We note that the green triangles in Fig.~\ref{fig:Eu_fe_GC-massive-gal} represent [Eu/Ca] measurements from \citet{Sakari2015}, rather than [Eu/Si], as Si abundances were not available for these clusters.

We use Figure\,\ref{fig:Eu_fe_GC-MW-dwarf} and Figure\,\ref{fig:Eu_fe_GC-massive-gal} to visually classify the GCs according to their proximity to the accreted and in-situ relations. Since these relations were calibrated using MW stellar populations, the resulting classifications for other galaxies should be regarded as empirical extensions of the MW framework. 
A GC is classified as ambiguous if its [Eu/Si] measurement error bar intersects both the accreted and in-situ trend lines. We also apply our classification criteria to the M31 outer-halo GCs from Sakari et al. (2015). The resulting classifications are listed in Table 1 alongside those from Monty et al. (2024). While the classifications agree for three of the five clusters, we classify PA17 and H23 as ambiguous because their [Eu/Si] uncertainties overlap both the in-situ and accreted relations.
The complete list of explored GCs is presented in Table\,\ref{tab:lg_gc_eus_fe}.

\begin{table*}
\caption{Iron and [Eu/Si] abundances for LG GCs. }
\label{tab:lg_gc_eus_fe}
\begin{tabular}{lcccclll}
\hline
\hline
\multicolumn{8}{c}{\citet{Larsen2022}}\\
\hline
ID & \ [Fe/H] & \ $\sigma_{Fe}^{*}$ & [Eu/Si] & [Eu/Si] error & in/ex-situ & Literature classifications & Source\\
\hline
NGC0104 & -0.735 & 0.019 & -0.151 & 0.067 & In-situ & In-situ (M-D)$^{**}$ & M25\\
&&&&&&in-situ&M24\\

NGC0362 & -1.073 & 0.019 & 0.470 & 0.069 & Accreted & Accreted (G-E)& M25\\
&&&&&&accreted&M24\\

NGC6254 & -1.485 & 0.018 & -0.091 & 0.103 & In-situ & LKH & M26\\
&&&&&&in-situ&M24\\

NGC6388 & -0.537 & 0.025 & -0.273 & 0.093 & In-situ & In-situ (M-B)& M25\\
&&&&&&most likely in-situ&M24\\

NGC6752 & -1.704 & 0.014 & -0.018 & 0.097 & In-situ & In-situ (M-D)& M25\\
&&&&&&in-situ&M24\\

NGC7078 & -2.293 & 0.020 & 0.060 & 0.158 & Accreted & Accreted (H99)/ In-situ (M-D)& M25\\
&&&&&&conflicting&M24\\

NGC7099 & -2.243 & 0.020 & -0.124 & 0.138 & Accreted & Accreted (G-E)& M25\\
&&&&&&in-situ&M24\\

M31 B006-G058 & -0.527 & 0.018 & -0.334 & 0.096 & In-situ & ... & ...\\
M31 B058-G119  & -0.985 & 0.023 & 0.011 & 0.084 & In-situ & ... & ...\\
M31 B163-G217  & -0.132 & 0.028 & -0.401 & 0.122 & In-situ & ... & ...\\
M31 B171-G222  & -0.217 & 0.024 & -0.451 & 0.129 & In-situ & ... & ...\\
M31 B225-G280  & -0.342 & 0.026 & -0.376 & 0.104 & In-situ & ... & ...\\
M31 B358-G219 & -2.199 & 0.023 & -0.147 & 0.321 & Ambiguous & ... & ...\\
M33 R12  & -0.861 & 0.023 & 0.287 & 0.248 & Accreted & prograde (In-situ/Accreted?) & A25\\
M33 CBF28 & -1.128 & 0.019 & 0.476 & 0.115 & Accreted & prograde (Accreted) & A25\\
M33 HM33B & -1.218 & 0.068 & 0.295 & 0.533 & Ambiguous & retrograde (In-situ/Accreted?) & L22 \\
&&&&&&&A25\\
N185 FJJ-III & -1.754 & 0.031 & -0.024 & 0.582 & Ambiguous & ... & ...\\ 
N185 FJJ-VIII & -1.749 & 0.026 & -0.033 & 0.452 & Ambiguous & ... & ...\\
N205 HubbleI & -1.410 & 0.025 & 0.440 & 0.193 & Accreted & ... & ...\\
N205 HubbleII & -1.348 & 0.023 & -0.203 & 0.235 & In-situ & ... & ...\\
Fornax 4 & -1.237 & 0.019 & 0.129 & 0.171 & In-situ & ... & ...\\
Fornax 5 & -2.058 & 0.027 & -0.355 & 0.699 & Ambiguous & ...& ...\\
N6822 SC6 & -1.689 & 0.025 & 0.511 & 0.285 & Accreted & ... & ...\\
N6822 SC7 & -1.130 & 0.019 & 0.155 & 0.174 & In-situ & ... & ...\\
N6822 HVII & -1.666 & 0.035 & -0.306 & 0.348 & In-situ & ... & ...\\ 
WLM GC & -1.849 & 0.029 & -0.105 & 0.413 & Ambiguous & ... & ...\\ 
N2403 F46 & -1.707 & 0.027 & 0.186 & 0.381 & Ambiguous & ... & ...\\
\hline
\hline
\multicolumn{8}{c}{\citet{Gvozdenko2024}}\\
\hline
Sextans~A GC-1 &-2.140 & 0.040 & -0.400 & +0.584$^*$ & In-situ & ... & ...\\ 

\hline
\hline
\multicolumn{8}{c}{\citet{Sakari2015}}\\
\hline
ID & \ [Fe/H] & \ $\sigma_{Fe}$ & [Eu/Ca] & [Eu/Ca] error & in/ex-situ & Literature classifications & Source\\
\hline
PA17 & -0.93 & 0.03 & 0.32 & 0.36 & Ambiguous & Accreted & M24\\
H23  & -1.12 & 0.02 & 0.12 & 0.25 & Ambiguous & In-situ & M24\\
H10  & -1.36 & 0.02 & 0.50 & 0.20 & Accreted & Accreted & M24\\
PA53 & -1.64 & 0.03 & 0.49 & 0.25 & Accreted &Accreted & M24\\
PA56 & -1.73 & 0.03 & 0.72 & 0.16 & Accreted & Accreted & M24\\
\hline
\end{tabular}

\begin{minipage}{\textwidth}
\vspace{2mm}
\footnotesize
Ambiguous - We cannot distinguish between the in-situ and accreted relations defined by \citet{Monty2024} \\
Sources: M25 = \cite{Massari2025}, M24 = \cite{Monty2024}, A25 = \cite{Anguiano2025}, L22 = \cite{Larsen2022},  
H24 = \cite{Hwang2014}.
\\
M25 \& M26 nomenclature: M-D = Main Disc (in-situ, disc-like orbits); M-B = Main Bulge (in-situ, tightly bound inner orbits); G-E = Gaia–Enceladus (accreted); LKH = Low-energy-Kraken-Heracles accretion event;
H99 = Helmi streams (accreted; \citealt{Helmi1999}). Where two associations are listed, the assignment is ambiguous. \\
$^{*}$ The $\sigma_{\mathrm{Fe}}$ values for the GCs are taken from \citet{Larsen2022} and correspond to the larger of the two quoted uncertainties provided in the data catalogue. \\
$^{**}$ Upper limit.\\
\end{minipage}
\end{table*}

\section{Discussion}
\label{sec:disc}

\subsection{The Milky Way}
For the 7 MW GCs which have integrated-light [Eu/Si] measurements \citep{Larsen2022}, four of these targets are consistent with being identified as in-situ (NGC~104, NGC~6254, NGC~6388, and NGC~6752), while two of the GCs are classified as accreted (NGC~0362 and NGC~7078) and only one GC (NGC~7099) has an error bar that crosses both trend lines of the relation (i.e., ambiguous).

All of these findings are consistent with the available literature. In particular, \cite{Massari2025} uses complete kinematic information, metallicity, and homogeneous age estimates to classify these clusters and associate them with specific progenitors.
NGC~104, NGC~6388, and NGC~6752 are all associated with the Main Progenitor population. Specifically, NGC~104 and NGC~6752 are identified as Main-Disc (M-D) clusters, while NGC~6388 is classified as a Main-Bulge (M-B) cluster. In contrast, NGC~362 and NGC~7099 are associated with the Gaia–Enceladus progenitor and are therefore classified as accreted clusters.
NGC~6254 is assigned to the low-energy (low-E) group in previous classifications. However, the recent analysis of \citet{Massari2026} identifies it as a member of the Low-energy-Kraken-Heracles (LKH) accretion event, which is thought to have occurred approximately 1.8 Gyr before the Gaia--Sausage--Enceladus merger.
Although these clusters define a relatively coherent Age–Metallicity Relation (AMR) with a high normalisation, suggesting the remnants of a potentially massive and as yet unidentified accretion event, the population is likely heterogeneous and its origin remains uncertain.
The classification of NGC~7078 is less clear and depends on the adopted distance scale. Using distances from \citet{Baumgardt2021}, the cluster is tentatively associated with the Helmi Streams progenitor \citep{Helmi1999}, whereas with \citet{Harris1996} distances it is instead classified as an in-situ Main-Disc cluster. The Helmi Streams system is generally linked to a lower-mass progenitor characterised by a comparatively low-normalisation AMR, and several of its proposed GC associations remain tentative. Consequently, the origin of NGC~7078 remains ambiguous, even though [Eu/Si] would suggest an accreted GC. 

We also compare our classifications with those of \citet{Monty2024}. The classifications are generally in agreement, with some differences at the metal-poor end, where the in-situ and accreted relations converge and become more difficult to distinguish. In particular, we classify NGC~7078 and NGC~7099 as accreted, whereas \citet{Monty2024} classify NGC~7078 as conflicting and NGC~7099 as in-situ. These differences highlight the difficulty of distinguishing the two populations at low metallicities, where the [Eu/Si] diagnostic becomes less clear.

\subsection{The M31 globular clusters}
\label{sec:m31}

\begin{figure*}
    \centering
    \includegraphics[width=12cm]{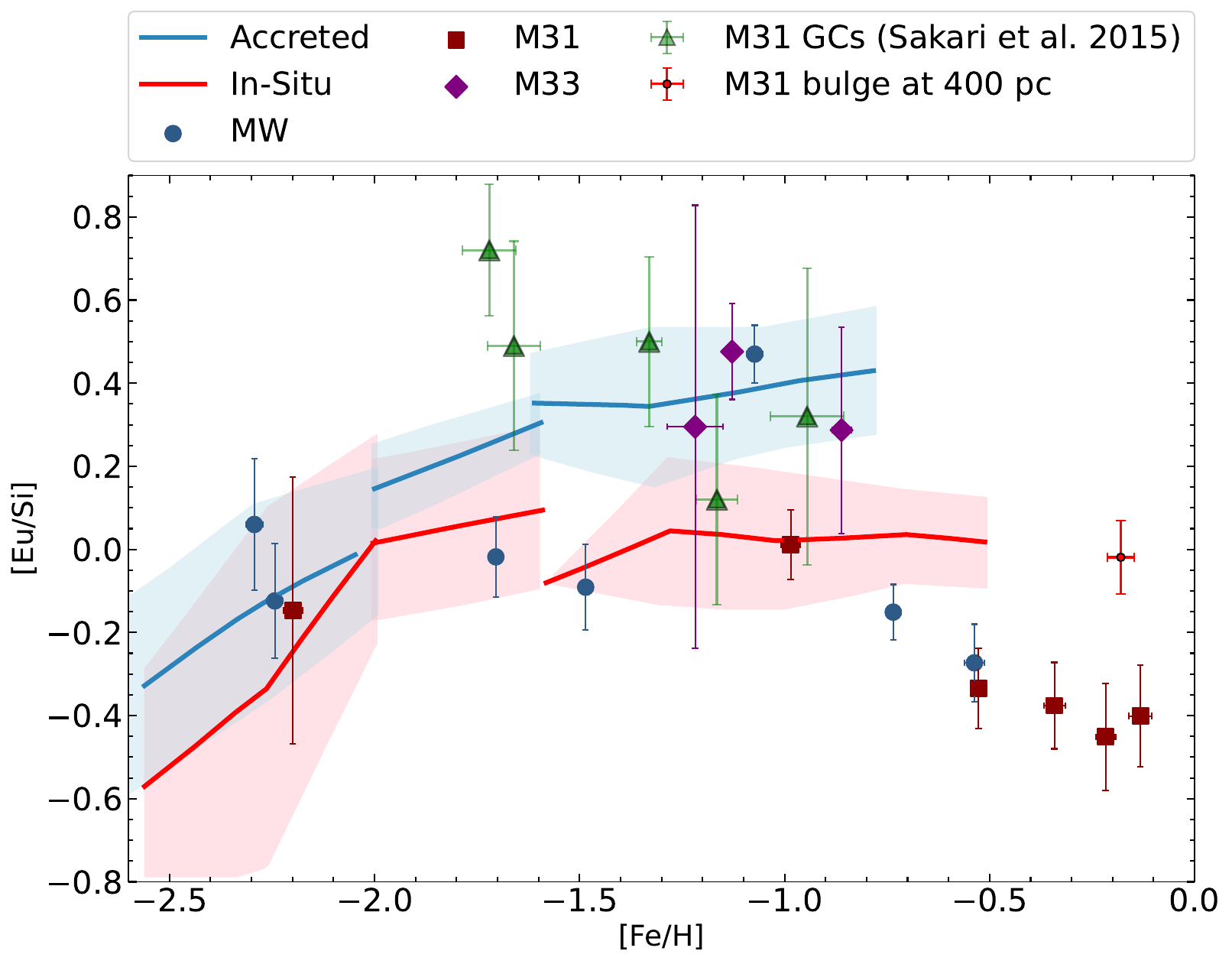}
    \caption{[Eu/Si] plotted against [Fe/H] for the massive galaxies GC systems data only, namely MW (circles), M31 (squares), and M33 (diamonds) \citep{Larsen2022}. Green triangles show the measurements from \citet{Sakari2015}  for which [Eu/Ca] is plotted instead of [Eu/Si], as Si abundances were not measured. The small red point is our bulge value for [Eu/Si]. The two lines are the same as with Figure\,\ref{fig:Eu_fe_GC-MW-dwarf}.}
    \label{fig:Eu_fe_GC-massive-gal}
\end{figure*}

In Figure\,\ref{fig:Eu_fe_GC-massive-gal}, we show the \cite{Larsen2022} data for M31, and include data for five "outer halo" GCs from \citet{Sakari2015}, where here Ca is used as a proxy for Si. 
We also include the integrated-light value for the M31 bulge stars at 400~pc\footnote{Computed as the average between 200 and 600~pc from the bulge centre, and adopted here as a representative value for the bulge as a whole.} from \citet{LaBarbera2026} including newly measured abundances outlined in Section~~\ref{sec:data}. 
There are six M31 GCs for which the Eu and Si abundances were derived in \cite{Larsen2022}. Five of these GCs,  M31~B058-G119, M31~B006-G058, M31~B225-G280, M31~B163-G217, and M31~B171-G222, occupy positions in Figure~\ref{fig:Eu_fe_GC-massive-gal} that are consistent with an in-situ origin. In contrast, the [Eu/Si] value of M31~358-219 does not provide a clear indication of its formation origin.
Again, we see that all five of the metal-rich M31 GCs lie at or below the extrapolation of the in-situ sequence defined by the MW stars and further extended by the M31 bulge value, while the outer halo GCs have high [Eu/Si] which is consistent with an accreted origin (a point already made by \citealt{Monty2024}). Furthermore, we notice that the bulge value of M31 lies on the direct extension of the 'in-situ' trend line defined for the MW stars. Interestingly, the metal-rich M31 GCs have a lower ratio of [Eu/Si] than seen in the M31 bulge stars at 400~pc. While speculative, this might be indicative of the relative formation timescales of the two systems in that the GCs could have formed before the bulk of the bulge stars. In this scenario, the bulge would have continued evolving chemically, producing additional Eu relative to Si and therefore increasing the [Eu/Si] ratio.
Alternatively, this behaviour may reflect a limitation of applying the original trend lines from \citet{Monty2024}, which were derived from MW GCs. This could indicate either differences in the chemical evolution of the M31 bulge or a distinct signature of the M31 GC system. However, the MW GCs NGC~6388 and NGC~0104, which lie in the $-0.8 < \mathrm{[Fe/H]} < -0.5$ range, follow the same trend as the M31 bulge GCs, favouring the interpretation that this behaviour is characteristic of GCs associated with the bulges of massive galaxies.

\begin{figure}
    \centering
    \includegraphics[width=9.5cm]{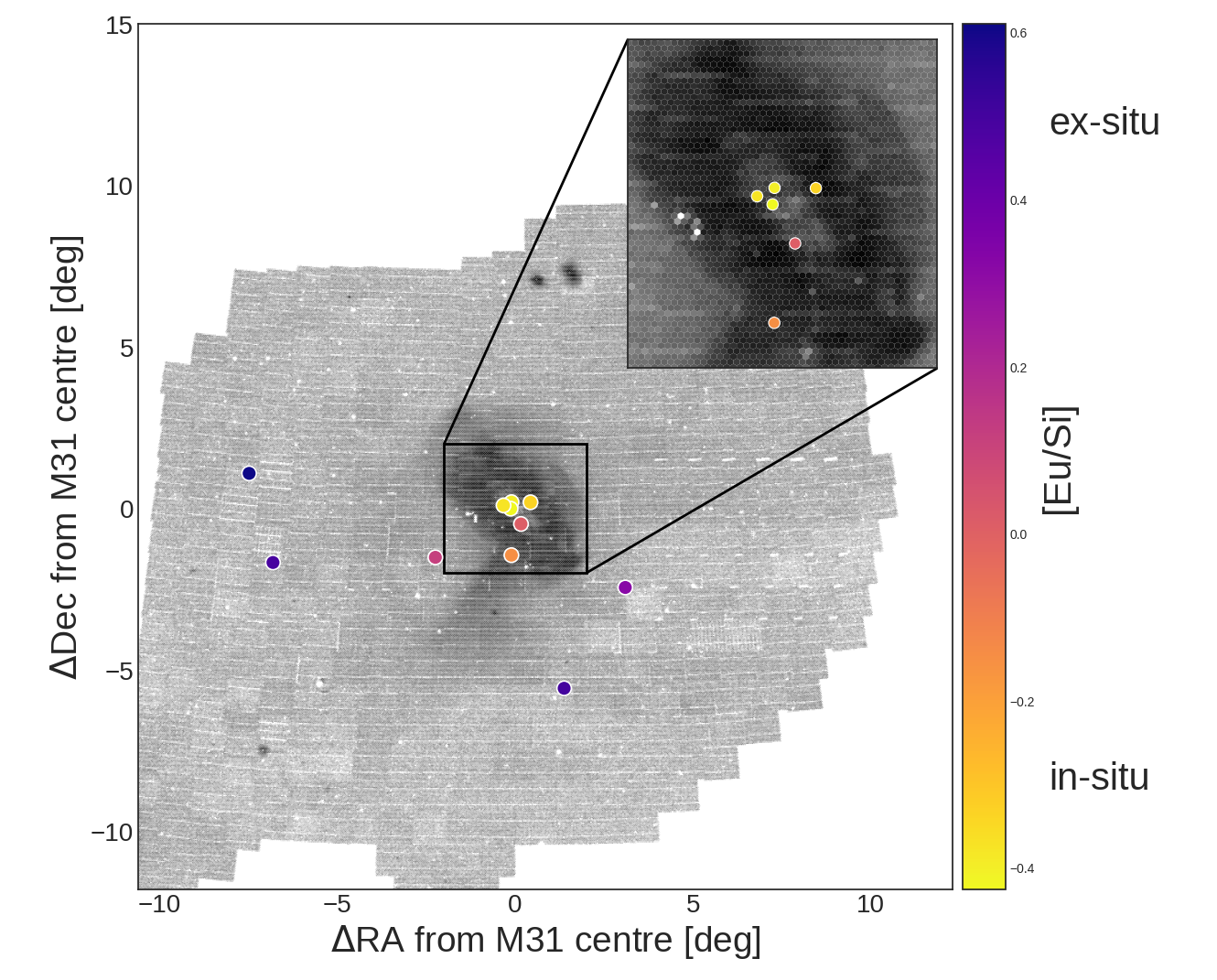}
    \caption{Locations of M31 (ours and \citealt{Sakari2015}) GCs overplotted on the PAndAS survey data \citep{McConnachie2009, McConnachie2018} for $0.0 < V-I < 2.0$. The PANDAS photometry has been thinned by a factor of 5 for display purposes. The GCs are colour coded by [Eu/Si]. 
    The ‘in-situ’ and ‘ex-situ’ labels indicate the tentative classifications based primarily on [Eu/Si], rather than on projected position. The projected spatial distribution is shown for context and should not be interpreted as an independent indicator of formation origin.
    In the case of the outer halo GCs of \citet{Sakari2015} Ca is used as a proxy for Si, since Si was not measured by these authors. The 'ex-situ' and 'in-situ' labels are indicative; the ratio of [Eu/Si] is not expected to be constant for either grouping for \\
    Fe/H $ <-2$.}
    \label{fig:spatialM31}
\end{figure}

We show the projected spatial distribution of the M31 GCs compared to the PAndAS survey photometry \citep{McConnachie2009, McConnachie2018} in Figure\,\ref{fig:spatialM31}, where we select stars in the colour range $0.0 < V-I < 2.0$ in order to primarily isolate the older RGB population.
The GCs from both this study and \citet{Sakari2015}  are colour coded by [Eu/Si].
The spatial distribution of the M31 GCs provides complementary context for their chemical abundance patterns. The low-[Eu/Si] GCs that we identify as consistent with an in-situ origin are located at relatively small projected radii, whereas the outer-halo GCs with high [Eu/Si] are consistent with an accreted origin. However, projected galactocentric distance alone cannot establish the formation origin of an individual GC, as both in-situ and accreted populations may coexist at a given radius. Studies of M31 GC substructures \citep{Mackey2019, Lewis2023} provide additional information on the possible accretion histories of individual clusters. The classifications adopted from \citet{Monty2024}, which are based on the chemical abundances measured by \citet{Sakari2015}, do not always agree with those of \citep{Mackey2019}, highlighting the fact that chemical and substructure-based classifications do not necessarily identify the same accretion history. None of the six M31 GCs in our spectroscopic sample is identified as a potential member of the Dulais Structure by \citet{Lewis2023}, or as belonging to a substructure in \citet{Mackey2019}. Nevertheless, the absence of an identified substructure association does not rule out an accreted origin, particularly for ancient accretion events whose debris may no longer be readily distinguishable. We therefore regard the spatial distribution as supporting context rather than independent evidence for our chemical classifications.
We note that \citet{McGill2025} recently found that M31 GCs with red HBs are associated with morphological substructures (in projection), irrespective of their metallicity. However, GCs associated with these substructures exhibit a range of HB morphologies. This suggests a possible link between red HBs and accretion events. Unfortunately, none of these GCs are in our spectroscopic samples.

\begin{figure*}
    
    \centering
    \includegraphics[width=13cm]{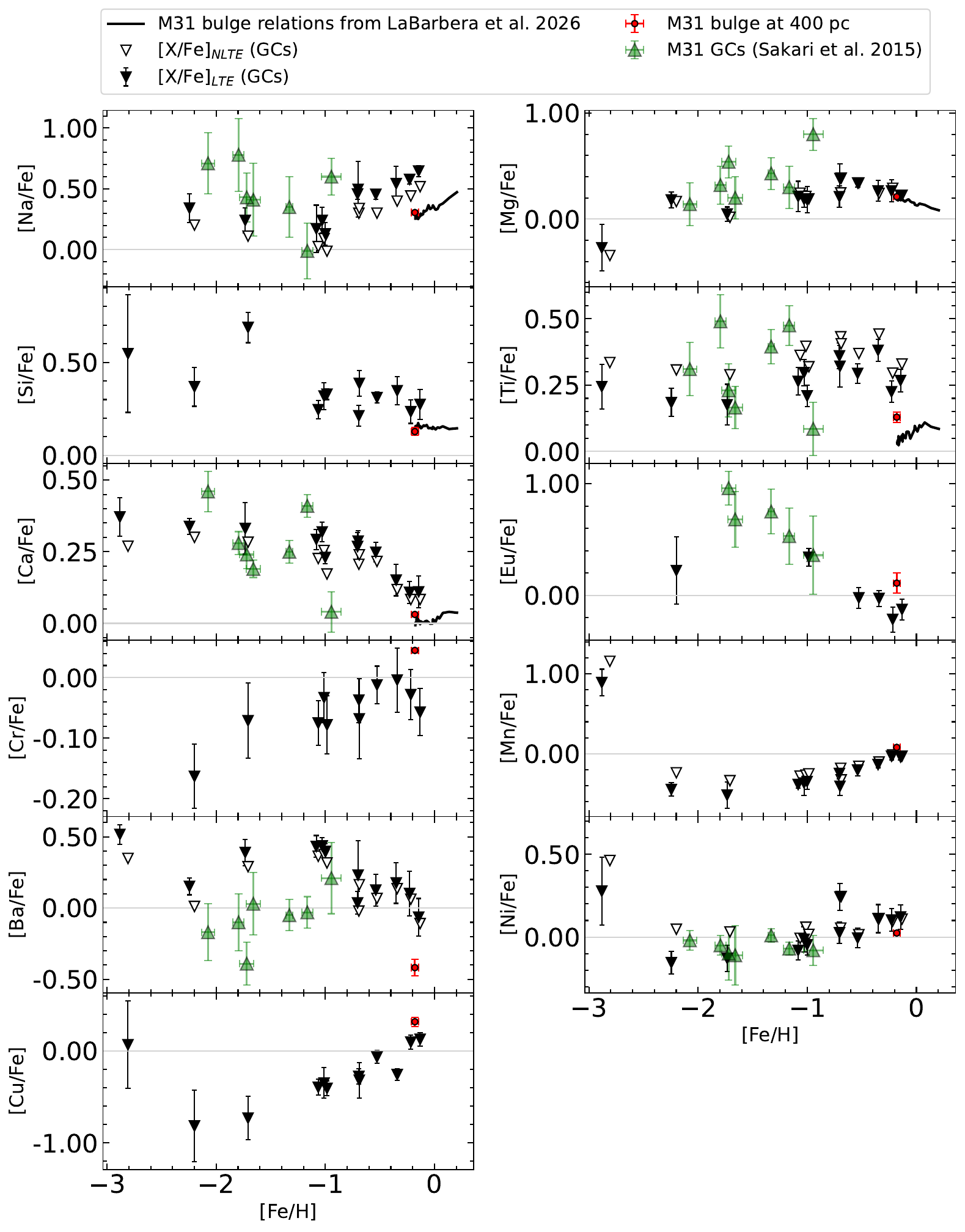}
    \caption{Abundance trends seen between Na, Mg, Si, Ti, Ca, Eu, Cr, Mn, Ba, Ni, and Cu vs Fe abundance ratio. The black data points are shown for the GCs data from \citet{Larsen2022}, and green for Na, Mg, Ti, Ca, Eu, Ba, and Ni; the GCs data are from \citet{Sakari2015}. For Na, Ca, Ti, Ba, Mn, and Ni two sets of data is shown: LTE (black symbols) and NLTE (white symbols) values both from \citet{Larsen2022}. The solid line represents the relations seen in \citet{LaBarbera2026} and the red square is the summed spectrum for the bulge in the radial range of 400 pc. The number of GCs shown varies between panels according to the availability of abundance measurements for each element. The plotted sample includes all available measurements from \citet{Larsen2022} and \citet{Sakari2015}.}
    \label{fig:M31_elem_ALLpanels}
\end{figure*}

We now compare the chemistry of the M31 GCs to that of the M31 bulge stars.
\cite{LaBarbera2026} provide a detailed investigation into the chemical abundance patterns within the M31 bulge, offering crucial insights into its formation and evolutionary history. 
By analysing the radial distribution of various elemental
ratios, the study uncovered a unique chemical signature, including strong central enhancements of Na, and an overall abundance pattern that strikingly resembles those found in the most massive early-type galaxies.
Figure\,\ref{fig:M31_elem_ALLpanels} highlights the trends of the abundance ratios of various elements versus [Fe/H]. For  Na, Ca, Ti, Ba, Mn, and Ni, we plot two sets of abundance values for \citet{Larsen2022}, LTE and NLTE. The SSP responses to individual abundance variations in \citet{LaBarbera2026} are computed with models from \citet{Conroy2018}, which are based on LTE calculations for all elements.
We note that, as shown in \citet{LaBarbera2026}, the M31 bulge exhibits no significant metallicity gradient beyond $\rm R \sim 200$~pc. Accordingly, our [Fe/H] estimate for the M31 stacked spectrum ($\sim -0.2$~dex; red points in Figure~\ref{fig:M31_elem_ALLpanels}) is fully consistent with the lowest metallicities reported by \citet{LaBarbera2026} (solid lines in the figure). We also note that the abundance ratios for the M31 stacked spectrum are derived here exclusively through full-spectral fitting (see the last paragraph of Section\,\ref{sec:data}), whereas LB26 combined different fitting techniques, including full-spectral and line-index fitting. This methodological difference accounts for the small offsets in [X/Fe] between the M31 stacked spectrum and the radial profiles from \citet{LaBarbera2026} (Figure~\ref{fig:M31_elem_ALLpanels}). In particular, as already discussed in \citet{LaBarbera2026}, full-spectral fitting tends to yield systematically higher [Ti/Fe] estimates than line-index fitting.       

Overall, the majority of the trends and abundance-ratio values seen for the M31 bulge and the  M31 GCs in \citet{Larsen2022} are in remarkable agreement; for example,  Si, Mg and Ti show similar [X/Fe] ratios for metallicities above $\sim$–1.0, a behaviour that is also followed by the M31 bulge. Ca shows a significant decline in the GCs for [Fe/H]$>-1.0$, with the most metallic GCs reaching [Ca/Fe]$\sim0.1$, which approach the bulge values. The Na measurements for the GCs show a strong increase at [Fe/H]$>-1.0$ reflecting a similar trend in the bulge stars.
[Eu/Fe] - measured at 400 pc in the M31 bulge - looks remarkably similar to the values determined for the most metal-rich M31 GCs.
For the heavier elements: Cr, Ba, Cu, Mn, and Ni, we again see remarkable agreement between the bulge and metal-rich GCs, with the possible exception of Ba. The fact that we see Ba somewhat suppressed in the bulge, while Eu is enhanced when compared to the metal-rich GCs, might suggest that \textit{r}-process sites are favoured over the s-process in the M31 central regions. 
Additionally this plot also highlights the comparison between \citet{Larsen2022} and \citet{Sakari2015} abundances. Even though the latter analysis results, which are LTE only, are more scattered and have larger error bars, they can be used for Eu and Ca (as mentioned above, can substitute Si) in this diagnostic. Namely, for Eu the [Eu/Fe] trend appears to be in agreement between the two studies, as well as for the [Ca/Fe] the trend is the same, only with the larger scatter for \citet{Sakari2015}. Therefore, the earlier comparison of the two analyses can be used without any corrections. To compare the \citet{Sakari2015} [Eu/Ca] values with \citet{Larsen2022} [Eu/Si] measurements, we calibrate the relation between the two ratios in Appendix\,\ref{app:euca_eusi}.

We also note in passing the intriguing behaviour of [Si/Mg] in the M31 bulge data. Si remains essentially constant with metallicity (radius), while Mg decreases with increasing metallicity (radius) implying an increasing ratio of [Si/Mg] towards the very central bulge regions. This divergence of [Si/Mg] -- both alpha-capture elements -- is an expected characteristic of the imprint of increasingly massive stellar progenitors where Si production is favoured over Mg \citep{Elham2026}.

In summary, with the possible exception of Ba, the element abundance ratios between the M31 GCs and the bulge at similar [Fe/H] are generally consistent within the measurement uncertainties, although some systematic differences are apparent.
This similarity suggests a close connection between the metal-rich GCs and the bulge stars, and reinforces the picture that the inner metal-rich M31 GCs are an in-situ population. Furthermore, we see a hint that the M31 bulge abundances fall systematically (slightly) below the GCs in all the $\alpha$-elements, Na, and Ba, but above the GCs in Eu. These small differences in Si and Eu combine to produce a more pronounced difference in [Eu/Si]. While the differences in [Ti/Fe] and [Cr/Fe] are not significant given the uncertainties (with the Ti comparison also potentially affected by differences in the treatment of LTE effects). Taken at face value, this potentially indicates that the GCs formed on shorter time-scales than the majority of the central bulge stars.

\subsection{M33 globular clusters}
\cite{Larsen2022} studied 8 M33 GCs, and for three the values of Eu were derived. Of the three clusters,
M33 R12 and M33 CBF28 appear unambiguously accreted in [Eu/Si], while M33 HM33B lies in the ambiguous region, taking into account the relatively large uncertainties on this measurement.
However, \cite{Larsen2022} already suggested that M33 HM33B might be an accreted GC -- based on its relatively low [$\alpha$/Fe] abundance -- and we do not dispute this assertion. 

These three clusters were also studied in \cite{Anguiano2025}, who identify a population of GCs on retrograde orbits in M33. They found that M33 HM33B is likely on a retrograde orbit, while both M33 R12 and M33 CBF28 (CB28 in their nomenclature) have prograde orbits. Retrograde orbits are a strong indicator of past accretion events, while prograde orbits plausibly entertain both accreted and in-situ scenarios. Thus, M33 HM33B is a candidate for an accreted GC, primarily based on its kinematics, although its chemical classification is uncertain given the large uncertainties in its [Eu/Si] measurement. 
the in-/ex-situ nature of M33~R12 and M33~CBF28 also remains unclear.

\subsection{Other galaxies}
\label{Othergalaxies}

There is very little work on the in-situ / ex-situ classifications of the GCs in the rest of the galaxies in our sample. The GCs of the lower mass galaxies (Fornax, NGC~205, NGC~185, WLM and Sextans~A) in our sample seem to show a mix of in-situ and accreted characteristics based on [Eu/Si]. That is, they may appear in-situ or accreted relative to their current dwarf hosts, but those dwarf hosts themselves perhaps resemble the kinds of galaxies that may go to build up the accreted halo populations of more massive galaxies. 

In the case of WLM and Sextans~A, these dwarfs are sufficiently isolated such that there is no obvious merger channel for them to have accreted GCs. In addition, each galaxy has a single GC with relatively small (projected) galactocentric radii (at 1 and 2 kpc for WLM and Sextans~A, respectively) \citep{Hodge1999, Beasley2019}. We classify both of these GCs as in-situ based on [Eu/Si].

The Fornax dSph is a satellite of the MW that has likely undergone several pericentric passages \citep{Fritz2018}. 
It is known to host six GCs, five of which are old, while the sixth is young. These clusters lie within $\sim$1.7 kpc galactocentric radius. However, given that Fornax is likely embedded in a significant dark matter halo, it is not clear why the GC orbits have not decayed to the centre via dynamical friction. One solution is that a cored dark matter halo would prevent the GCs from further infall via "core-stalling" \citet{Cole2012}. The authors note, however, that if the dark matter halo is only "mildly cuspy", then the GCs would likely have to have come from further out  (r $>3-5$ kpc) which might then suggest that they could be an accreted population. We classify the two Fornax GCs as in-situ based on [Eu/Si].

NGC6822 is a dwarf irregular galaxy ($\sim1\times10^{8}M_\odot$; \citealt{Weldrake2003}) with 11 known GCs \citep{Hwang2011, Huxor2013, Veljanoski2015, Howell2026}. The GCs are aligned in a linear fashion with the major axis of the spheroid, which is misaligned with the major axis of the gas disc, and exhibit mild rotation. 
This alignment therefore does not necessarily indicate an association with the gas disc or an in-situ disc population \citep{Veljanoski2015}. However, four of these clusters have relatively large half-light radii (7.5--14 pc), and galactocentric radii ranging from 1.5 to 11 kpc, possibly indicating an accreted origin for a subset. The GCs in our sample (SC6, SC7 and Hubble VII) all lie within 3.0 kpc of the galaxy centre,  with Hubble VII essentially coincident with the centre. 
We classify SC7 and Hubble VII as in-situ, while SC6 lies in the accreted region in [Eu/Si]. However, the origin of SC7 is less clear when considering other properties. \citet{Larsen2022}, \citet{Howell2026} and \citet{McGill2026} highlight its high luminosity and mass, high ellipticity, and low [$\alpha$/Fe], which have been interpreted as suggestive of an accreted origin. This is therefore at odds with our [Eu/Si]-based classification of SC7 and highlights the possibility that different diagnostics may provide contrasting indications of its origin.

NGC~185 and NGC~205 are dwarf spheroidals/ellipticals orbiting M31. Of the two, NGC 205 appears the more morphologically disturbed system, suggesting an interaction with M31 \citep{Sato1986, Choi2002}. In addition, NGC~205 appears to have GCs with a range of ages \citep{Sharina2006}, indicating that it may have formed star clusters up until recently. Interestingly, for the "unperturbed" NGC~185, we classify both GCs FJJ-III and FJJ-VIII as in-situ, while in the case of NGC 205, Hubble I appears accreted while Hubble II lies firmly in the in-situ region based on [Eu/Si].

NGC2403 is classified as a late-type Scd similar in stellar mass to M33 ($\sim6\times10^{9}M_\odot$; \citealt{Corbelli2003}). It has at least 7 known GCs \citep{Battistini1984, Davidge2007, Forbes2022, Larsen2025}, with an expected population of order $\sim50$ GCs \citep{Larsen2025}. 
\citet{Forbes2022} suggest, based on a sample of 7 GCs,  that there may be a pronounced AMR amongst the GC population. Our only GC in this galaxy, F46, is also in the \citet{Forbes2022} sample, and they identify it as an old halo cluster, i.e., likely an in-situ GC. Based on [Eu/Si] while we classify this cluster as "ambiguous",  given that the uncertainty of the [Eu/Si] measurement is large and it overlaps with both the accreted and in-situ branches. 

\subsection{Formation origins of the GC systems}

GCs in low-mass galaxies are expected to resemble accreted clusters in massive systems; however, accreted dwarf progenitors may differ in their stellar populations from present-day surviving dwarf galaxies.
It is furthermore a debatable subject what would count as accreted; if the merger happened at earlier times, the originally accreted GC might resemble the in-situ population quite closely. 
Such a case could be an early merger with a relatively massive progenitor, as has been proposed for the Milky Way (the Kraken/Heracles event; \citealt{Kruijssen2019, Kruijssen2020, Horta2021}), whose associated GCs are difficult to distinguish chemically from in-situ clusters (Horta et al. 2020). M31 also appears to have experienced multiple accretion events (\citealt{Mackey2019}; see Section\,\ref{sec:m31}), including a major merger proposed by \citep{DSouza2018}.

The [Eu/Si]–[Fe/H] tracks adopted from Monty et al. (2024) are empirically defined using Milky Way stellar populations with different likely formation origins, rather than being derived from chemical evolution models. They provide an observational reference for distinguishing populations with different enrichment histories. 
Although the [Eu/$\alpha$] diagnostic is calibrated using the MW and dwarf galaxies, applying it to massive galaxies assumes broadly similar chemical evolution pathways. This assumption should be treated with caution because massive galaxies experience different star-formation histories and enrichment timescales.
In particular, the assembly history of M31 likely differs substantially from that of the Milky Way \citep{DSouza2018, Mackey2019}, which may affect the location and extent of its in-situ and accreted populations in the [Eu/Si]–[Fe/H] plane. Consequently, we use the Milky Way relations as a reference for interpreting the chemical abundances of M31 GCs, rather than assuming that they provide an exact or universal separation between in-situ and accreted populations.

With the above caveat in mind, as illustrated in Figure~\ref{fig:Eu_fe_GC-MW-dwarf} and Figure~\ref{fig:Eu_fe_GC-massive-gal}, the GCs identified as accreted are preferentially located at higher [Eu/Si] values. The clearest separation between the in-situ and accreted populations occurs at intermediate metallicities ($-2 < [\mathrm{Fe/H}] < -1$), where both populations can be identified. At lower metallicities, the two relations converge, making it increasingly difficult to distinguish between the two origins using [Eu/Si]. In addition, while we identify a mixture of in-situ and accreted GCs in intermediate-mass dwarf galaxies, the GCs in the lower-mass dwarf galaxies in our sample are predominantly classified as in-situ. Taken together, these results suggest that accreted GCs in massive galaxies were contributed, at least in part, by intermediate-mass dwarf progenitors, whose GCs occupy the intermediate-metallicity regime. This is also consistent with the metallicities of the metal-poor GC populations in massive galaxies, which are generally thought to be predominantly accreted (e.g. \citealt{Beasley2018}). However, our sample is relatively small, and the convergence of the [Eu/Si] relations at low metallicities means that we may be missing accreted GCs at the lowest metallicities, if such a population exists.
The elevated [Eu/Si] values of these accreted GCs are consistent with progenitors that experienced extended star-formation histories, allowing sufficient time for significant \textit{r}-process enrichment (e.g. Eu), and hence they may be systematically younger than in-situ counterparts formed in more rapid, efficient episodes of star formation. In contrast, the minimal scatter observed among metal-rich clusters suggests a uniform formation pathway, likely characterised by short formation timescales and highly efficient star formation inside massive haloes.

The general trend we see is that the GCs with the higher metallicity contain smaller scatter in [Eu/$\alpha$] compared to the metal-poor end of the sample. This could be an effect correlated with the uncertainties, which are generally larger for the GCs with lower iron abundances. However, some of the scatter may be intrinsic, as fewer stellar progenitors contribute to the abundances as [Fe/H] decreases. For a few elements in particular, e.g., Mg and Na, this may also be due to intrinsic abundance spreads within the GCs becoming more prominent at lower metallicities (see, e.g. \citealt{Larsen2022}).

\section{Conclusions}
\label{sec:concs}
We have explored the feasibility of chemical tagging in extragalactic systems using a high-quality, homogeneous dataset of
integrated-light abundances for 21 Local Group GCs (28 including the MW GCs with [Eu/Si] measurements).
We primarily focus on the [Eu/Si] relation for a subset of  28 Local Group GCs to investigate their formation origins. Based on the inferred trends of \citet{Monty2024}, we classify GC systems in the MW, M31, M33, NGC~185, NGC~205, Fornax, NGC~6822, WLM, NGC~2403, and Sextans~A into accreted and in-situ born populations. We further verify these classifications against the literature, where available, finding general agreement with independent constraints from kinematics, metallicity, and homogeneous age estimates, and our [Eu/Si]-based classification.

We also examine chemical abundance patterns in the M31 bulge and compare them with those of the centrally located GCs, providing key constraints on the formation and evolutionary history of the bulge.
The chemical properties of the metal-rich M31 GCs appear closely linked to the bulge trends reported by \citet{LaBarbera2026} for Na, Ca, Mg, and Si, and by extension to those of massive early-type galaxies.
The relative location of the metal-rich GCs with respect to the bulge in different chemical dimensions may further suggest that they formed on shorter star-formation timescales, potentially preceding the bulk of bulge evolution. In particular, the GCs exhibit slightly enhanced $\alpha$-element and Na abundances relative to the bulge, while showing lower Eu and Ba abundance ratios.

Overall, [Eu/Si]-based classification of GC origins proves effective across the Local Group, revealing:
\begin{itemize}
    \item[--] metal-rich M31 GCs are consistent with an in-situ origin based on their [Eu/Si] behaviour;
    \item[--] evidence for mixed formation histories in the NGC~6822 and NGC~205 GC systems;
    \item[--] accreted GCs in massive galaxies likely originate, at least in part, from intermediate-mass, intermediate-metallicity progenitors. This inference is supported by the mixture of in-situ and accreted GCs found in intermediate-mass dwarf galaxies and the metallicity range over which the [Eu/Si] diagnostic most clearly separates the two populations. The associated extended star-formation histories would allow Eu enrichment to build up over time, potentially resulting in comparatively younger GCs;
    \item[--] the minimal scatter observed among metal-rich GCs suggests a broadly uniform formation pathway, characterised by short formation timescales and highly efficient star formation.
\end{itemize}

These results demonstrate that [Eu/Si] provides a promising, physically motivated chemical diagnostic for extragalactic archaeology, offering new insight into the formation pathways of galaxies beyond the MW.

\section*{Acknowledgements}
We thank the anonymous referee for a careful and critical reading of the manuscript.
MAB and AV acknowledge funding from the Spanish Science Ministry via the "MATRYOSHKA" project PID2024-162088NB-I00. KF acknowledges funding from the European Union’s Horizon Europe research and innovation programme under the Marie Sk\l{}odowska-Curie grant agreement No 101103830.
AG was supported by the STFC through consolidated grant ST/X001075/1.
MAB and AV acknowledge financial support from the Spanish Agencia Estatal de Investigación (AEI) through the Severo Ochoa Centre of Excellence accreditation awarded to the Instituto de Astrofísica de Canarias, grant CEX2025-001609-S, funded by MICIU/AEI/10.13039/501100011033.
AG and MAB thank the ExGal-Twin project, funded by the European Union (Widening Participation, ExGal-Twin, GA 101158446) for support. Views and opinions expressed are those of the author(s) only and do not necessarily reflect those of the European Union. Neither the European Union nor the granting authority can be held responsible for them.

\section*{Data Availability}

The data underlying this article are available in the cited literature, primarily \citet{Larsen2022}, \citet{Gvozdenko2024}, and \citet{LaBarbera2026}. The additional abundance measurements derived in this work for the M31 bulge are provided within the article and its appendix. Further data products are available from the corresponding author upon reasonable request.



\bibliographystyle{mnras}
\bibliography{bibl}



\appendix
\section{M31 bulge measurements}
\label{apx:M31bulge}

\citet{LaBarbera2026} measure elemental abundance ratios in the bulge of M31 using high signal-to-noise long-slit optical spectroscopy from OSIRIS/GTC, spatially binned in radius to build integrated-light spectra of unresolved stellar populations. They infer [X/Fe] and [Fe/H] by comparing these spectra to the EMILES stellar population synthesis models \citep{Vazdekis2015}, allowing age, metallicity, and individual element abundances to vary. \citet{LaBarbera2026} employ and cross-check three complementary approaches -- full-spectrum fitting, simultaneous fitting of multiple spectral indices, and classical line-strength analysis -- calibrated using mock spectra and SDSS stacked galaxy spectra to validate systematics. Element sensitivities are extracted through response-function-based modelling of absorption features (e.g. Mg, Na, CN, Ca lines) based on theoretical response functions \citep{Conroy2018}. The method relies on comparisons between observed spectra of the M31 bulge and stacked early-type galaxy (ETG) spectra from the Sloan Digital Sky Survey (SDSS), and a suite of stellar population synthesis (SSP) models incorporating variations in individual elemental abundances.

Long-slit spectroscopy of the M31 bulge along its major axis was obtained using the OSIRIS instrument on the Gran Telescopio Canarias (GTC). 
For comparison, stacked ETG spectra from SDSS are also employed. These data provide the observational basis for constraining radial trends in metallicity and chemical composition.
The analysis is performed using SSP models drawn from the E-MILES, CvD18, and Padova00 libraries. These empirical SSP spectra are modified using theoretical response functions to account for variations in individual elemental abundances. The underlying stellar libraries predominantly follow MW abundance patterns, which are approximately scaled-solar at solar and super-solar metallicities, with the exception of oxygen, for which metallicity-dependent corrections are applied due to its observed decrease with increasing metallicity in the solar neighbourhood.

In the present work, elemental abundance ratios are derived using full spectral fitting (FSF), in which the entire observed spectrum is fitted to SSP models by minimising a $\chi^2$ statistic.  The fitting procedure yields abundance variations, $\delta[X/\mathrm{H}]$, relative to the baseline SSP abundance pattern. While the empirical SSP models are approximately scaled-solar, iron is allowed to vary independently to increase flexibility. The iron metallicity is therefore computed as
\begin{equation}
[\mathrm{Fe}/\mathrm{H}] = [Z/\mathrm{H}] + \delta[\mathrm{Fe}/\mathrm{H}],
\end{equation}
where $[Z/\mathrm{H}]$ is the total metallicity of the SSP models. Individual abundance ratios are then obtained as
\begin{equation}
[X/\mathrm{Fe}] = \delta[X/\mathrm{H}] - \delta[\mathrm{Fe}/\mathrm{H}].
\end{equation}

\begin{table}
\caption{LaBarbera PC bulge abundance measurements (value $\pm$ error).}
\label{tab:labarbera_bulge}
\begin{tabular}{lcc}
\hline
Element & [X/Fe] & error \\
\hline

$[\mathrm{Mg}/\mathrm{Fe}]$ & 0.211 & 0.009 \\
$[\mathrm{Na}/\mathrm{Fe}]$ & 0.305 & 0.018 \\
$[\mathrm{C}/\mathrm{Fe}]$ & 0.157 & 0.016 \\
$[\mathrm{Ca}/\mathrm{Fe}]$ & 0.031 & 0.006 \\
$[\mathrm{N}/\mathrm{Fe}]$ & 0.231 & 0.020 \\
$[\mathrm{Ti}/\mathrm{Fe}]$ & 0.129 & 0.019 \\
$[\mathrm{Si}/\mathrm{Fe}]$ & 0.128 & 0.021 \\
$[\mathrm{O}/\mathrm{Fe}]$ & 0.242 & 0.045 \\
$[\mathrm{Cr}/\mathrm{Fe}]$ & 0.045 & 0.004 \\
$[\mathrm{Mn}/\mathrm{Fe}]$ & 0.080 & 0.015 \\
$[\mathrm{Ba}/\mathrm{Fe}]$ & -0.418 & 0.057 \\
$[\mathrm{Ni}/\mathrm{Fe}]$ & 0.024 & 0.012 \\
$[\mathrm{Co}/\mathrm{Fe}]$ & 0.167 & 0.011 \\
$[\mathrm{Eu}/\mathrm{Fe}]$ & 0.109 & 0.090 \\
$[\mathrm{Sr}/\mathrm{Fe}]$ & 0.139 & 0.043 \\
$[\mathrm{Cu}/\mathrm{Fe}]$ & 0.318 & 0.050 \\
\hline
[Fe/H] & -0.180 & 0.032 \\
\hline
\end{tabular}
\end{table}

\section{Labelled plot}
\label{apx:plot_IDs}
The same plot as Figure\,\ref{fig:Eu_fe_GC-MW-dwarf} and Figure\,\ref{fig:Eu_fe_GC-massive-gal} combined, but with labels indicating the IDs of the globular cluster systems.
\begin{figure*}
    \centering
    \includegraphics[width=14cm]{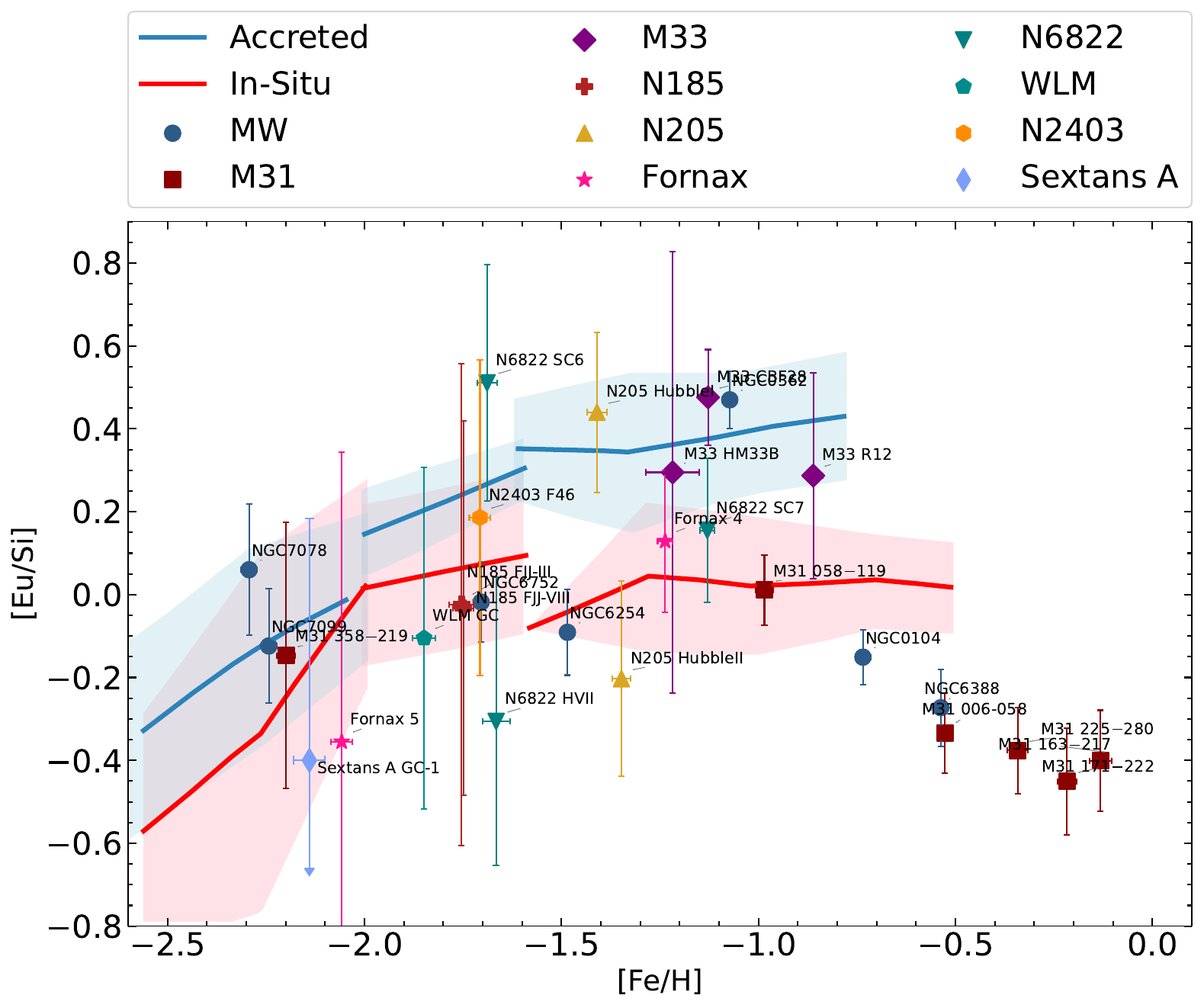}
    \caption{Same as Figure\,\ref{fig:Eu_fe_GC-MW-dwarf} and Figure\,\ref{fig:Eu_fe_GC-massive-gal} combined but with ID labels from \citet{Larsen2022}.}
    \label{fig:Eu_fe_GC_withNames}
\end{figure*}

\section{Relation between [Eu/Ca] and [Eu/Si] in M31 globular clusters}
\label{app:euca_eusi}

\begin{figure}
    \centering
    \includegraphics[width=\linewidth]{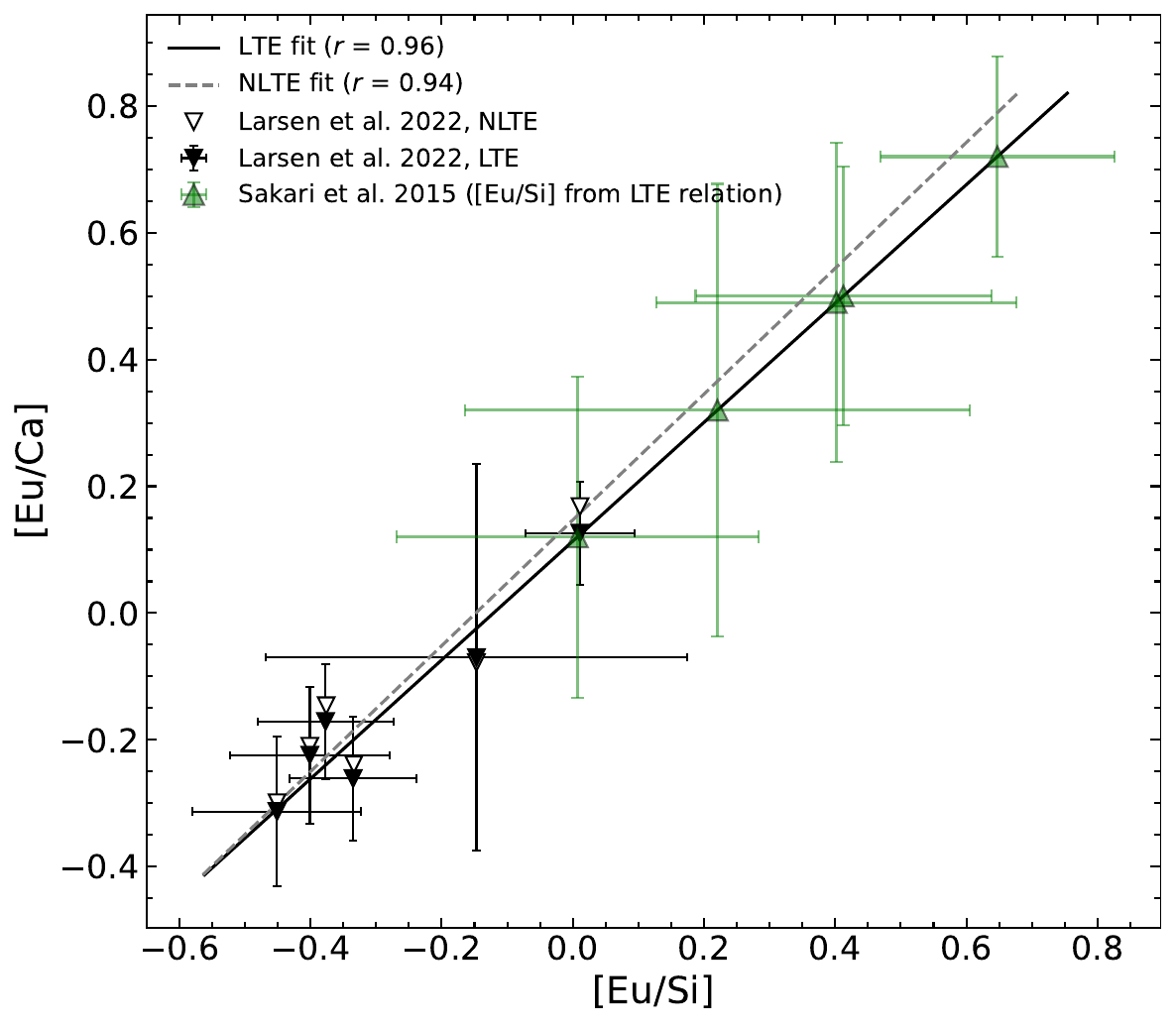}
    \caption{[Eu/Ca] versus [Eu/Si] for the six M31 GCs of \citet{Larsen2022}, where each symbol is one cluster with both ratios measured from the same spectrum. Black-filled triangles show the LTE abundances and white-filled triangles the NLTE abundances. Ca has NLTE corrections, hence there are two different values for [Eu/Ca], while Si only has LTE values. The solid black and dashed grey lines are linear fits of [Eu/Si] as a function of [Eu/Ca] to the LTE ($r=0.96$) and NLTE ($r=0.94$) measurements, respectively. Green triangles show the five M31 GCs of \citet{Sakari2015} with Eu measurements. Their [Eu/Ca] values are as measured (LTE), and their [Eu/Si] values are inferred from the \citet{Larsen2022} LTE relation. The horizontal error bars on these points combine the propagated [Eu/Ca] uncertainty with the rms scatter about that relation (0.06\,dex). Clusters with Eu, Si or Ca flagged as unmeasured in \citet{Larsen2022} are omitted.}
\label{fig:euca_eusi}
\end{figure}

Part of our M31 analysis relies on the [Eu/Ca] measurements of \citet{Sakari2015}, whereas our discussion is based mainly on [Eu/Si]. To place these measurements on a common scale, we use the six M31 GCs of \citet{Larsen2022} for which both [Eu/Ca], in LTE and NLTE, and [Eu/Si], in LTE only, are available. Fig.~\ref{fig:euca_eusi} shows [Eu/Ca] against [Eu/Si] for these clusters, together with linear fits to the LTE and NLTE measurements. The two ratios are strongly correlated, with Pearson coefficients of $r=0.96$ (LTE) and $r=0.94$ (NLTE) and an rms scatter about the LTE relation of 0.06\,dex. The LTE relation is
\begin{equation}
{\rm [Eu/Si]} = 1.07\,{\rm [Eu/Ca]} - 0.12,
\end{equation}
so with a slope close to unity, [Eu/Si] is lower than [Eu/Ca] by about 0.1\,dex. Since the \citet{Sakari2015} measurements were derived in LTE, we use this relation to convert their [Eu/Ca] values to [Eu/Si], and show the results as green triangles in Fig.~\ref{fig:euca_eusi}. The converted values range from $+0.01$ to $+0.65$\,dex.

\bsp	
\label{lastpage}
\end{document}